\documentclass[preprint,12pt]{elsarticle}
\usepackage[utf8x]{inputenc}
\usepackage{graphicx}
\usepackage{amsmath}
\usepackage{amsbsy}
\usepackage{MnSymbol}

\newcommand{\MCA}[1]{\left\langle #1 \right\rangle_{\rm MC}}

\begin{document}

\begin{frontmatter}
\title{Improved Green's Function Measurement for Hybridization Expansion Quantum Monte Carlo}
\author[UA,FZU]{\corref{cor1}Pavel Augustinsk\'y}
\ead{august@fzu.cz}
\author[FZU]{Jan Kune\v{s}}
\ead{kunes@fzu.cz}
\address[UA]{Theoretical Physics III, Center for Electronic Correlations and Magnetism, 
Institute of Physics, University of Augsburg, D-86135, Augsburg, Germany}
\address[FZU]{Institute of Physics, Czech Academy of Sciences, Cukrovarnick\'a 10/112, 16200, Prague, Czech Republic}
\cortext[cor1]{Corresponding author}
\begin{abstract}
\noindent
We present an algorithm for measurement of
the Green's function in the hybridization expansion continuous-time quantum
Monte-Carlo based on continuous estimators. Compared to the standard method, 
the present algorithm has similar or better accuracy with improvement 
notable especially at low perturbation orders and high frequencies. 
The resulting statistical noise is weakly correlated so high-accuracy data
for the numerical analytical continuation can be produced.
\end{abstract}
\end{frontmatter}

\section{Introduction}
The Monte-Carlo simulations are among the most widely used techniques for investigation
of quantum impurity models. The progress of the dynamical mean-field theory 
\cite{dmft,ldadmftb,ldadmftc} in the 
past decade, which allowed material specific studies of multi-band Hubbard models, 
is to a large extent governed by the availability
of reliable and fast solvers for multi-orbital impurity problems. 
New generation quantum Monte-Carlo (QMC) techniques based on the stochastic sampling
of perturbative expansions \cite{Werner06a,Rubtsov05,Haule07,Gull11} quickly became 
the standard of quantum impurity simulations.

Several QMC algorithms are currently used in this context differing by the type 
of expansion and symmetry of the impurity Hamiltonian.
As the simulation of impurities with arbitrary interaction \cite{Werner06b,Lauchli09} is
computationally demanding, specialized algorithms that can handle a simplified
density-density interaction at much lower computational cost still play an important
role.

In this article we describe an improved technique for measurement of Green's functions
in the strong-coupling CT-QMC algorithm (also called CT-HYB) for density-density
interactions.
The development was motivated by poor performance of the standard measurement
technique
in systems with strongly imbalanced perturbation orders for different orbitals. Such 
a situation is common in materials containing orbitals with fluctuating occupation
coupled by interaction to filled or empty orbitals,
e.g. the $t_{2g}$ orbitals in nickelates \cite{Deng12}, $e_g$ orbitals in early transition metal 
oxides \cite{Held01,Nekrasov05}, $t_{2g}$ orbitals in the low-spin state
of LaCoO$_3$ \cite{Krapek12}, or Hubbard model with crystal-field splitting \cite{Werner07,Kunes11}.
Although the filled orbitals experience little fluctuations themselves and their influence on the
active
(partially filled) orbitals is well approximated by a static potential, it is their
Green's functions that may become prohibitively noisy. In some cases one can simply
remove the filled orbitals from the effective model, e.g. early transition metals
are commonly studied
with $t_{2g}$-only models \cite{Pavarini04}. 
Often, however, 
some orbitals became filled/empty only in a certain part of the phase diagram 
and then 
they have to be kept
in the model, e.g. to study spin/orbital ordering, or
because one is interested in their dynamics, which may exhibit non-trivial 
dynamical renormalization \cite{Kunes07,Deng12}.

The paper is organized as follows. In section \ref{section_CTHYB} we review the basic of the 
CT-HYB algorithm and the standard measurement technique. In section \ref{section_measurementG} 
we discuss on a general level the relationship between the expansions for the 
partition function and the Green's function. The new measurement algorithm is described in section
\ref{section_Improved_measurement} with details and implementation notes summarized in section 
\ref{section_implementation_notes}.
Finally, in section \ref{section_results}
we provide examples and performance tests.

\section{CT-HYB}
\label{section_CTHYB}
At this point, we briefly review the CT-HYB algorithm introduced in \cite{Werner06a,Werner06b} 
(for details see comprehensive review \cite{Gull11}).

We wish to study the dynamics of the $d$-electrons in the Anderson impurity model 
with \emph{density-density} interaction 
described by the Hamiltonian
\begin{equation}
 H = \sum_{\alpha k} \epsilon_{\alpha {k}} n_{\alpha k} 
   + \sum_\alpha n_\alpha \Big( E^d_\alpha  + \sum_{\beta} U_{\alpha\beta} n_\beta \Big)
   + \sum_{\alpha k} \Big( V_{\alpha {k}} c_{\alpha { k}} d^\dagger_\alpha + h.c. \Big)
   \label{H_imp}
\end{equation}
consisting of three parts $H = H_{bath} + H_{loc} + H_{hyb}$. 
Bath fermionic operators are denoted by $c^{\phantom\dagger}_{\alpha { k}},\ c^\dagger_{\alpha { k}}$ while
$d^{\phantom\dagger}_\alpha,\ d^\dagger_\alpha$ stand for the impurity operators and 
$n_\alpha = d^\dagger_\alpha d^{\phantom\dagger}_\alpha$ are the corresponding occupation number operators. 
The flavor index $\alpha$ represents both the spin and the orbital quantum number.
Our aim is to evaluate the Green's function 
\begin{equation}
 G_{\alpha}(\tau', \tau'') = - \langle T d^{\phantom \dagger}_\alpha(\tau'') d^\dagger_\alpha(\tau') \rangle.
\label{G_imp}
\end{equation}
We assume the bath and the Green's function to be diagonal in the flavor indices
throughout the paper but generalization to non-diagonal baths is possible.

The CT-HYB algorithm is based on the expansion of the partition function in the hybridization strength
\begin{align}
  Z &= {\rm Tr}\,[{\rm e}^{-\beta H}] = {\rm Tr}\left[ {\rm e}^{-\beta (H_{bath} + H_{loc})} T 
  {\rm e}^{-\int_0^\beta H_{hyb}(\tau) d \tau}  \right] = \nonumber \\
 &= \sum_{k=0}^\infty (-1)^k \int_0^\beta d \tau_1 \ldots \int_{\tau_{k-1}}^\beta d \tau_k 
{\rm Tr} \big[ {\rm e}^{-\beta (H_{bath} + H_{loc})} 
H_{hyb}(\tau_k)\ldots H_{hyb}(\tau_1) \big].
\label{Z_expansion}
\end{align}
For $H_{loc}$ diagonal in the occupation number basis, expansion (\ref{Z_expansion})
can be viewed as a sum over so called \emph{segment configurations} ($Z$-configurations). 
In a schematic way, it can be written as
\begin{equation}
  Z = Z_{bath}\sumint d \boldsymbol{\tau} \, Z(\boldsymbol{\tau}) = 
  Z_{bath} \sumint d \boldsymbol{\tau} \, Z_{loc}(\boldsymbol{\tau}) \, Z_{hyb}(\boldsymbol{\tau}),
\end{equation}
The partition function of the bath $Z_{bath}$ presents only an 
irrelevant multiplication factor.
The abbreviation $\boldsymbol{\tau} \equiv \{\{ \tau^s_{i},\tau^e_{i} \}_\alpha\}$
is used here to describe all the integration variables, that is
a set of start-times 
and end-times for all flavors. We stress that we use symbol $\boldsymbol{\tau}$
to describe an \emph{arbitrary} set of numbers $\tau^s_{i}$ and $\tau^e_{i}$ from interval $(0,\beta)$. 
If $\boldsymbol{\tau}$ does not represent an allowed $Z$-configuration, $Z(\boldsymbol{\tau})$ is zero.
Otherwise, we have
\begin{align}
 Z_{loc} &= \exp\left( -\sum n_\alpha E^d_\alpha - \sum_{\alpha < \beta} U_{\alpha\beta} o_{\alpha\beta} \right),
\nonumber \\
Z_{hyb} &= \prod_\alpha \det F_\alpha(\tau^e_i-\tau^s_j).
\end{align}
where $n_\alpha$ and $o_{\alpha\beta}$ stand for the total length of segments of flavor 
$\alpha$ and their overlap with segments of flavor $\beta$, and $F_\alpha(\tau)$
is the hybridization function.
A random series of $Z$-configurations is generated
using the Metropolis importance sampling with 
the probability density $Z(\boldsymbol{\tau})$.
This series is then used to obtain various quantities of interest 
such as the Green's function (\ref{G_imp}). 
In the standard approach, from each $Z$-configuration with $K$ segments of flavor
$\alpha$ one obtains an estimator of the Green's function that consist of $K^2$ 
delta-functions
\begin{equation}
 G_{\alpha}(\tau',\tau'') = \MCA{\sum_{ij} M^\alpha_{ij}
 \delta(\tau'-\tau^s_i)\delta(\tau''-\tau^e_j)}.
\label{G_estimator_standard}
\end{equation}
Here, $M^\alpha_{ij}$ stands for the matrix element of the inverse 
hybridization matrix of flavor $\alpha$.
With this approach, however, difficulties are encountered when the mean perturbation order of the simulation 
is low. In such a case, the estimator (\ref{G_estimator_standard}) consists of only a few $\delta$-functions and 
a poor statistics results. Moreover, due to the discrete character of estimators, a significant statistical
noise is observed in $G_\alpha(i\omega_n)$, and even worse in the selfenergy, at higher frequencies.

Recently, a substantial progress has been achieved in fixing these problems.
The statistical noise in the selfenergy (and the two-particle Green's function) can be 
significantly suppressed when improved estimators based on the equation of motion 
are used \cite{Hafermann12}. Besides that, filtering out the stochastic
noise using orthogonal polynomial representation can be used to 
further improve the results \cite{Boehnke11}. Nevertheless, neither of these methods can fully avoid
the poor statistics at low perturbation orders. 

\section{Measurement of the Green's function}
\label{section_measurementG}
There is not a unique way to 
estimate the Green's function
from the random series of $Z$-configurations.
In fact, there is a substantial freedom in this procedure and expression 
(\ref{G_estimator_standard}) presents only one of the possibilities.
In this section, we review the general logic of the measurement 
and establish notation used in the rest of the paper.

Perturbation expansion of the Green's function can be, 
similarly to series (\ref{Z_expansion}), written as 
\begin{equation}
  G_\alpha(\tau', \tau'') = \frac 1 Z \sumint d \boldsymbol{\tau} \, 
  G_\alpha(\tau', \tau''; \boldsymbol{\tau})
  = \frac 1 Z \sumint d \boldsymbol{\tau} \, 
  G_{loc} (\tau', \tau''; \boldsymbol{\tau})
  G_{hyb} (\boldsymbol{\tau}).
\label{G_expansion}
\end{equation}
It can be represented as a sum over segment 
configurations with one pair of special start-time and end-time
at $\tau'$ and $\tau''$ ($G$-configurations).
In CT-HYB, series (\ref{G_expansion}) is integrated by the 
Monte-Carlo method with $Z(\boldsymbol{\tau})$ used as the 
probability density for the importance sampling.
Unless $\tau'=\tau''$, there are regions of the $\boldsymbol{\tau}$-space 
where $Z(\boldsymbol{\tau})$ is identically zero while 
$G_\alpha(\tau', \tau''; \boldsymbol{\tau})$ is not. Therefore, $Z(\boldsymbol{\tau})$
can \emph{not} be used to directly sample expansion (\ref{G_expansion}).
In the standard approach, to each
$G$-configuration $\{\tau', \tau'', \boldsymbol{\tau}_1\}$ 
we assign \emph{one} $Z$-configuration $\boldsymbol{\tau}_2$   
so that 
if $G_\alpha(\tau', \tau''; \boldsymbol{\tau}_1)$ is non-zero, 
$Z(\boldsymbol{\tau}_2)$ is non-zero as well. Then, once $\boldsymbol{\tau}_2$ is 
visited during the random walk, $G_\alpha(\tau', \tau''; \boldsymbol{\tau}_1)/Z(\boldsymbol{\tau}_2)$
is accumulated as the estimator for $G_\alpha(\tau', \tau'')$. 
Drawback of this approach is that 
we obtain information about $G_\alpha(\tau', \tau'')$ only 
when $\tau'$ and $\tau''$ correspond to one start-time and
one end-time in the visited $Z$-configuration.

The standard measurement can be generalized when to each $G_\alpha(\tau', \tau''; \boldsymbol{\tau}_1)$ 
we assign a \emph{set} of configurations $Z(\boldsymbol{\tau}_2)$
with some weight distribution $w(\tau',\tau'',\boldsymbol{\tau}_1,\boldsymbol{\tau}_2)$
and accumulate
\begin{equation}
w(\tau',\tau'',\boldsymbol{\tau}_1,\boldsymbol{\tau}_2)\,\frac{G_\alpha(\tau', \tau''; \boldsymbol{\tau}_1)}{Z(\boldsymbol{\tau}_2)} =
w(\tau',\tau'',\boldsymbol{\tau}_1,\boldsymbol{\tau}_2)\,\frac{G_{loc}(\tau', \tau'', \boldsymbol{\tau}_1) G_{hyb}(\boldsymbol{\tau}_1)}
{Z_{loc}(\boldsymbol{\tau}_2)Z_{hyb}(\boldsymbol{\tau}_2)}
\label{estimator_G_general}
\end{equation}
when \emph{any} $\boldsymbol{\tau}_2$ is visited. 
In principle, the weight function can be chosen arbitrarily as long as normalization condition 
\begin{equation}
\int d\boldsymbol{\tau}_2 \,w(\tau',\tau'',\boldsymbol{\tau}_1,\boldsymbol{\tau}_2) = 1
\label{normalization_condition_general} 
\end{equation}
is fulfilled for each $\tau',\tau''$ and $\boldsymbol{\tau}_1$.
However, for the sake of efficiency,
$w(\tau',\tau'',\boldsymbol{\tau}_1,\boldsymbol{\tau}_2)$ should be
non-zero only when it is cheap to evaluate ratio 
$G_\alpha(\tau', \tau''; \boldsymbol{\tau}_1)/Z(\boldsymbol{\tau}_2)$
in equation (\ref{estimator_G_general}).

\section{Improved measurement}
\label{section_Improved_measurement}
Here, we introduce a choice of the weight function 
$w(\tau',\tau'',\boldsymbol{\tau}_1,\boldsymbol{\tau}_2)$
which constitutes the improved measurement.
For a given $Z$-configuration, we assign a non-zero weight to a set of $G$-configurations 
generated by 
removing of a hybridization line followed by a shift of the lone 
$d_\alpha^{\phantom\dagger}(\tau_j''),\ d_\alpha^\dagger(\tau_j')$ operators to all 
positions consistent with the remaining segments. 
This way, 
from any $Z$-configuration (apart from the zeroth order of the perturbation 
theory) we obtain a non-zero contribution to $G_\alpha(\tau', \tau'')$
for all values of arguments so the estimator of the Green's function is a 
continuous function of the imaginary time.

In the following, we distinguish two types of $G$-configurations. 
Those where $\tau'$ can be shifted to $\tau''$ without crossing a segment boundary
will be called \emph{connected}. Remaining ones
will be called \emph{separated}. The Green's function 
can be split into two terms corresponding to sums of series (\ref{G_expansion}) 
over the connected and the separated configurations as 
\begin{equation}
 G_\alpha(\tau', \tau'') = G_\alpha^C(\tau', \tau'') + G_\alpha^S(\tau', \tau'').
\end{equation}
Their evaluation is discussed in next two subsections.

\subsection{Separated configurations}
\label{subsection_separated}
\begin{figure}
\centering
  \includegraphics{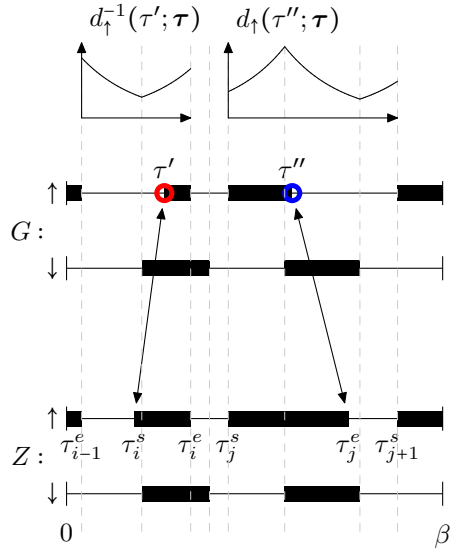} 
 \caption{Relation between the separated $G$-configuration (top) and a $Z$-configuration 
 (bottom). When we remove the hybridization line connecting $\tau^s_i$ and $\tau^e_j$, 
 we can reach $G$-configurations with $\tau'' \in (\tau^e_{i-1},\tau^e_i)$ and 
  $\tau' \in (\tau^s_j,\tau^s_{j+1})$. The piecewise-exponential $\tau$-dependence of 
  functions $d_\uparrow(\tau;  \boldsymbol{\tau})$ and 
  $d_\uparrow^{-1}(\tau;  \boldsymbol{\tau})$ is schematically depicted in the 
  two graphs in the upper part of the figure. Regions with different atomic states are 
  separated by the vertical dashed lines.}
\label{figure_Z2G}
\end{figure}
First, we discuss the evaluation of $G_\alpha^S(\tau', \tau'')$ which is somewhat simpler.
In expression (\ref{estimator_G_general}), the ratio of the bath traces equals the element of the 
inverse hybridization matrix as in the standard formula (\ref{G_estimator_standard}).
The local traces differ by the position of one start-time and one end-time. 
At a fixed segment configuration, the local operators of flavor $\alpha$ ``feel'' 
an effective $\tau$-dependent potential
\begin{equation}
 E^d_\alpha(\tau;\boldsymbol{\tau}) = E^d_\alpha + \sum_{\beta\neq \alpha} U_{\alpha\beta} \, n_\beta(\tau;\boldsymbol{\tau})
\label{effective_atomic_energy}
\end{equation}
consisting of the site energy $E^d_\alpha$ and the interaction 
contribution due to all segments of other flavors. We define a function 
\begin{equation}
 d_\alpha(\tau; \boldsymbol{\tau}) = \exp\left(-\int_0^\tau d \sigma E^d_\alpha(\sigma;\boldsymbol{\tau})\right)
 \label{def_dtau}
\end{equation}
so that 
shifting of the creation operator from $\tau^s_i$ to $\tau'$ results 
in scaling the local trace by factor
${d_\alpha(\tau^s_i;\boldsymbol{\tau})}/{d_\alpha(\tau';\boldsymbol{\tau})}$.
When annihilation operator is moved from $\tau^e_j$ to $\tau''$, the local
trace is scaled by ${d_\alpha(\tau'';  \boldsymbol{\tau})}/{d_\alpha(\tau^e_j;\boldsymbol{\tau})}$.
The lower bound of the integral in equation (\ref{def_dtau}) can be chosen arbitrarily 
since it influences only an irrelevant multiplicative factor that cancels out.
The estimator for the separated Green's function then reads
\begin{equation}
G^S_\alpha(\tau', \tau'';\boldsymbol{\tau}) = 
\sum_{\substack{ ij \\ j \neq i, i+1 }} 
  M^\alpha_{ij} \frac{d_\alpha(\tau'';\boldsymbol{\tau})d_\alpha(\tau^s_i;\boldsymbol{\tau})}
                     {d_\alpha(\tau^e_j;\boldsymbol{\tau})d_\alpha(\tau';   \boldsymbol{\tau})}
  w(\tau', \tau'',\tau^s_i,\tau^e_j;\boldsymbol{\tau}). 
\label{estimator_G_separated_tau}
\end{equation}
The summation does not include neighboring pairs of start- and end-times 
since they contribute to the 
connected part of the Green's function. We switched from abstract notation 
for weight $w(\tau',\tau'',\boldsymbol{\tau}_1,\boldsymbol{\tau}_2)$ and we 
explicitly indicate relevant imaginary-time arguments. 
In this notation, the normalization condition (\ref{normalization_condition_general}) reads
\begin{equation}
\int_{\tau^e_{i-1}}^{\tau^e_i} d\tau^s_i \int_{\tau^s_j}^{\tau^s_{j+1}} d\tau^e_j \, 
w(\tau', \tau'',\tau^s_i,\tau^e_j;\boldsymbol{\tau}) = 1.
\label{normalization_condition_tau}
\end{equation}

So far, only the support of $w$ was specified while
its explicit functional form remains undetermined. 
In the following, we restrict ourselves
to a factorized form 
\begin{equation}
 w(\tau', \tau'',\tau^s_i,\tau^e_j;\boldsymbol{\tau}) = 
 w_c(\tau',\tau^s_i;\boldsymbol{\tau}) w_a(\tau'',\tau^e_j;\boldsymbol{\tau}).
\label{weight_factorization}
\end{equation}
The standard measurement formula (\ref{G_estimator_standard}) 
is restored with the $\delta$-function weight 
$w_c=\delta(\tau''-\tau^s_i)$, $w_a=\delta(\tau'-\tau^e_j)$. Since we want to
improve on this and use the entire accessible imaginary-time interval, the 
uniform weight 
$w_c=1/(\tau^e_i-\tau^e_{i-1} + \beta \theta(\tau^e_{i-1}-\tau^e_i))$, 
$w_a=1/(\tau^s_{j+1}-\tau^s_j+\beta \theta(\tau^s_{j}-\tau^s_{j+1}))$ 
appears like a natural choice. However, our numerical experiments showed 
that it yields strong noise 
especially for strongly interacting systems and low temperatures and it 
is always outperformed by
the ``normalization'' weight
\begin{align} 
 w_c(\tau^s_i, \tau';\boldsymbol{\tau}) &= d_\alpha^{-1}(\tau^s_i;\boldsymbol{\tau}) \left( \int_{\tau^e_{i-1}}^{\tau^e_i} 
d_\alpha^{-1}(\sigma;\boldsymbol{\tau}) d\sigma  \right)^{-1}\chi_{(\tau^e_{i-1},\tau^e_i)}(\tau'),
\nonumber \\
 w_a(\tau^e_j, \tau'';\boldsymbol{\tau}) &= d_\alpha(\tau^e_j;\boldsymbol{\tau}) 
 \left( \int_{\tau^s_j}^{\tau^s_{j+1}} d_\alpha(\sigma;\boldsymbol{\tau}) d\sigma \right)^{-1} \chi_{(\tau^s_j,\tau^s_{j+1})}(\tau'').
\label{normalization_weight_separated}
\end{align}
Here, $\chi_{(a,b)}(\tau)$ stands for the characteristic function of interval $(a,b)$.
With this weight, each $\delta$-function of the standard formula (\ref{G_estimator_standard}) is smeared
to a piecewise-exponential function with \emph{unit} integral regardless of the initial 
position of $\tau^s_i$ and $\tau^e_j$. This property makes the algorithm stable at 
any regime.

In principle, we can store estimator (\ref{estimator_G_separated_tau}) in any basis during the simulation. 
Nevertheless, the imaginary time representation is inefficient because we must either 
introduce a large discretization error or use a fine imaginary-time grid which requires 
many numerically costly evaluations of expression (\ref{estimator_G_separated_tau}).
Despite its clear advantages, the recently proposed orthogonal polynomial representation for the 
Green's function \cite{Boehnke11} is not well suited for our purpose either and we use the 
Matsubara basis.
Its main advantage lies in the factorization property 
$\exp(i\omega_n (\tau'-\tau''))=\exp(i\omega_n \tau')\exp(-i\omega_n \tau'')$ that reduces the computational 
complexity of the measurement. Besides that, it is 
simpler to Fourier transform a piecewise-exponential function than calculate its Legendre 
polynomial representation. In Matsubara frequencies with 
normalization weight (\ref{normalization_weight_separated})
we obtain 
\begin{align}
 G^S_\alpha(i\omega_n) &= 
 \frac 1 {\beta^2} \int_0^\beta\int_0^\beta d\tau' d\tau'' \exp(i\omega_n (\tau''-\tau')) 
  \, G^S_\alpha(\tau', \tau'') \nonumber \\
 &= \left\langle \sum_{\substack{ ij \\ j \neq i, i+1 }}
M^\alpha_{ij} C^\alpha_i(i\omega_n) A^\alpha_j(i\omega_n) \right \rangle_{\rm MC}
\label{Gw_separated}
\end{align}
where functions $A^\alpha_j(i\omega_n)$ and $C^\alpha_j(i\omega_n)$ are given by 
\begin{align}
 A^\alpha_j(i\omega_n) &= \frac 1 \beta \int_{\tau^s_j}^{\tau^s_{j+1}} d\tau'' 
 \frac{d_\alpha(\tau'';  \boldsymbol{\tau})}{d_\alpha(\tau^e_j;\boldsymbol{\tau})}
  w_a(\tau^e_j, \tau'';\boldsymbol{\tau}) \exp(i\omega_n \tau'') 
  = \frac{\tilde{A}^\alpha_j(i\omega_n)}{\tilde{A}^\alpha_j(0_+)}, 
  \nonumber \\
 \tilde{A}^\alpha_j(i\omega_n) &= \frac 1 \beta \int_{\tau^s_j}^{\tau^s_{j+1}} d\tau'' 
 d_\alpha(\tau'';  \boldsymbol{\tau}) \exp(i\omega_n \tau''), 
 \nonumber \\
 C^\alpha_i(i\omega_n) &= \frac 1 \beta \int_{\tau^e_{i-1}}^{\tau^e_i} d\tau' \frac{d_\alpha(\tau^s_i;  \boldsymbol{\tau})}{d_\alpha(\tau';\boldsymbol{\tau})}
  w_c(\tau^s_i, \tau';\boldsymbol{\tau}) \exp(-i\omega_n \tau')
   = \frac{\tilde{C}^\alpha_j(i\omega_n)}{\tilde{C}^\alpha_j(0_+)},
   \nonumber\\
 \tilde{C}^\alpha_i(i\omega_n) &= \frac 1 \beta \int_{\tau^e_{i-1}}^{\tau^e_i}
 d\tau' d_\alpha^{-1}(\tau';\boldsymbol{\tau}) \exp(-i\omega_n \tau').   
\label{def_AC_omega}
\end{align}

In principle, the piecewise-exponential functions $d_\alpha(\tau';  \boldsymbol{\tau}),\ d_\alpha(\tau'';  \boldsymbol{\tau})$
could be modulated by a to some extent arbitrary function 
if a $\tau', \tau''$-dependent weight was chosen. However, the 
normalization condition (\ref{normalization_condition_tau}) does not 
allow to chose the modulation independently for each $\tau^s_i,\ \tau^e_j$ and this
makes the choice of a better weight non-trivial. 
It remains an open question whether one can choose it in a more clever way so that it will, for instance, 
simplify numerical evaluation of equation (\ref{def_AC_omega}).

\subsection{Connected configurations}
\begin{figure}
 \centering
 \includegraphics{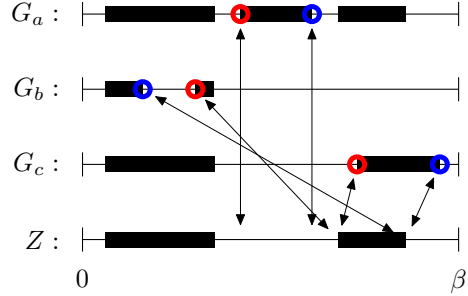}
 \caption{Different ways how a connected Green's function configuration can be reached from 
 partition function configuration $Z$. $G_a$ corresponds to insertion method. $G_b$ and $G_c$
 can be obtained by removing of a hybridization line and shifting the remaining operators.}
 \label{figure_Z2G_diag}
\end{figure}

There are two different simple ways to reach a 
connected $G$-configuration from a $Z$-configuration.
The first option is to 
remove a hybridization line spanning
a single segment or antisegment and shift $d_\alpha(\tau''),\ d^\dagger_\alpha(\tau')$
to a new position. 
The second option is to insert a $d_\alpha(\tau''),\ d^\dagger_\alpha(\tau')$ pair into
a $Z$-configuration. 
This approach is routinely used for estimation of the occupation numbers, that is
the equal time Green's function. It can be used for measurement of the 
full Green's function for Hamiltonians including the spin-flip 
terms \cite{Haule07}. But for density-density
interactions, the insertion measurement alone is not ergodic since
it can not access the separated $G$-configurations.
The two possibilities are shown in Figure \ref{figure_Z2G_diag} 
and we will discuss them separately.

\subsubsection{``Remove and shift'' approach}
\label{section_RS_approach}
From the two types of $G$-configurations $G_b$ and $G_c$ depicted in 
Figure \ref{figure_Z2G_diag}, we use a non-zero weight only for the later ones.
Then, estimator for $G^C_\alpha(i\omega_n)$ is derived analogously to
equation (\ref{estimator_G_separated_tau}). The only 
difference here is that the sum on the right hand side goes over complementary indices and
$\tau'$ and $\tau''$ must be properly ordered. In Matsubara frequencies we obtain
\begin{equation}
 G^C_\alpha(i\omega_n) = \MCA{ \sum_i 
  M^\alpha_{ii} P^\alpha_i(i\omega_n) +
  M^\alpha_{i+1,i} H^\alpha_i(i\omega_n) } 
\label{Gw_conn_RS}
\end{equation}
where 
\begin{align}
 P^\alpha_i(i\omega_n) &= \frac 1 {\beta^2} \int_{\tau^e_{i-1}}^{\tau^s_{i+1}} d\tau' \int_{\tau'}^{\tau^s_{i+1}} d\tau'' 
\frac{d_\alpha(\tau'';  \boldsymbol{\tau})}{d_\alpha(\tau^e_j;\boldsymbol{\tau})}
\frac{d_\alpha(\tau^s_i;  \boldsymbol{\tau})}{d_\alpha(\tau';\boldsymbol{\tau})}
  w(\tau^s_i,\tau^e_j, \tau',\tau'';\boldsymbol{\tau})
 \exp(i\omega_n (\tau''-\tau')), \nonumber \\
 H^\alpha_i(i\omega_n) &= -\frac 1 {\beta^2} \int_{\tau^s_i}^{\tau^e_{i+1}} d\tau'  \int_{\tau^s_i}^{\tau'} d\tau'' 
\frac{d_\alpha(\tau'';  \boldsymbol{\tau})}{d_\alpha(\tau^e_j;\boldsymbol{\tau})}
\frac{d_\alpha(\tau^s_i;  \boldsymbol{\tau})}{d_\alpha(\tau';\boldsymbol{\tau})}
 w(\tau^s_i,\tau^e_j, \tau',\tau'';\boldsymbol{\tau})
 \exp(i\omega_n (\tau''-\tau')).
\label{def_P_H_omega}
\end{align}
Again, we can ensure the normalization with a weight function
\begin{equation}
 w(\tau^s_i,\tau^e_j, \tau',\tau'';\boldsymbol{\tau}) 
= N\frac{d_\alpha(\tau^e_j;\boldsymbol{\tau})}{d_\alpha(\tau^s_i;  \boldsymbol{\tau})}
\label{normalization_weight_connected}
\end{equation}
where the multiplication factor $N$ is chosen so that condition 
(\ref{normalization_condition_general}) is fulfilled. 
In the definition of $P^\alpha_i(i\omega_n)$ we have 
\begin{equation}
 \frac 1 N = \int_{\tau^e_{i-1}}^{\tau^s_{i+1}} d\tau^s_i \int_{\tau^s_i}^{\tau^s_{i+1}} d\tau^e_j \, 
\frac{d_\alpha(\tau^e_j;\boldsymbol{\tau})}{d_\alpha(\tau^s_i;  \boldsymbol{\tau})}
\end{equation}
and analogously for $H^\alpha_i(i\omega_n)$.

Individual estimators given by equation (\ref{Gw_conn_RS}) 
do not have the exact $1/\omega$ asymptotic. Therefore, the stochastic
noise in the Green's function decays with the inverse frequency and it is correlated.
However, when apart from the Green's function we accumulate also the leading 
order coefficient in $G^C_\alpha(i\omega_n) \approx -i G_\alpha^\infty/\omega_n$, 
we can at the end of the simulation
evaluate the Green's function as $G^C_\alpha(i\omega_n)/G_\alpha^\infty$. 
After this correction, the noise decays with the inverse \emph{square} of
frequency and its correlations are \emph{suppressed}.

\subsubsection{``Insertion'' approach}
With the insertion approach, there is only one $Z$-configuration from which 
a particular $G$-configuration can be created. Therefore, as in the standard 
measurement, the weight is uniquely determined by normalization condition
(\ref{normalization_condition_general}). When there is at least one segment
of flavor $\alpha$, the estimator reads
\begin{align}
 G^C_\alpha(i\omega_n) = \MCA{
 \sum_i  \tilde{P}_i^\alpha(i\omega_n) +  \tilde{H}_i^\alpha(i\omega_n)}.
\end{align}
Unlike in equation (\ref{Gw_conn_RS}), elements of the inverse hybridization matrix are missing here and definitions of
$ \tilde{P}_i^\alpha(i\omega_n)$ and $\tilde{H}_i^\alpha(i\omega_n)$ cover shorter imaginary-time 
intervals (corresponding to only one segment or antisegment)
\begin{align}
  \tilde{P}_i^\alpha(i\omega_n)&= 
 \frac 1 {\beta^2} 
 \int_{\tau^s_i}^{\tau^e_i} d\tau' \int_{\tau^s_i}^{\tau'} d\tau'' 
\frac{d_\alpha(\tau'';  \boldsymbol{\tau})}{d_\alpha(\tau';\boldsymbol{\tau})}
 \exp(i\omega_n (\tau''-\tau')),
 \nonumber \\
 \tilde{H}_i^\alpha(i\omega_n) &= 
 -\frac 1 {\beta^2} 
 \int_{\tau^e_i}^{\tau^s_{i+1}} d\tau' \int_{\tau'}^{\tau^s_{i+1}} d\tau'' 
\frac{d_\alpha(\tau'';  \boldsymbol{\tau})}{d_\alpha(\tau';\boldsymbol{\tau})}
 \exp(i\omega_n (\tau''-\tau')).
\end{align}

Unlike with the RS method, we obtain
non-zero estimator $I_0^\alpha(i\omega_n)$ even from $Z$-configurations with 
no segments of flavor $\alpha$.
Then, there are only two possible states: the empty-line and the full-line. 
Given the configuration of other segments, probabilities of the two are given by 
$p_0 = 1/(1+d_\alpha(\beta;\boldsymbol{\tau})$ and 
$p_1 = d_\alpha(\beta;\boldsymbol{\tau})/(1+d_\alpha(\beta;\boldsymbol{\tau})$. 
We can use this and write
\begin{align}
  I_0^\alpha(i\omega_n)&= 
 \frac 1 {\beta^2} \frac{1}{1+d_\alpha(\beta;\boldsymbol{\tau})}
 \int_{0}^{\beta} d\tau' \int_{\tau'}^{\tau'+\beta} d\tau'' 
\frac{d_\alpha(\tau'';  \boldsymbol{\tau})}{d_\alpha(\tau';\boldsymbol{\tau})}
 \exp(i\omega_n (\tau''-\tau')).
 \label{def_I_0}
\end{align}
This estimator is accumulated \emph{regardless} of whether we reach 
the empty-line or the full-line state. 

With the insertion method, each estimator obeys the exact 
$G_\alpha(i\omega_n) \rightarrow 1/i\omega_n$ asymptotic
because the sum of lengths of all segments and antisegments 
equals to $\beta$ and estimator (\ref{def_I_0}) is normalized 
by the probability prefactor. Therefore, the stochastic noise
scales with $\omega^{-2}$ at high frequencies. 

\subsubsection{Comparison with RS method}
When the mean perturbation order is very low, the insertion 
method is more accurate then the RS method since it works 
directly from the zeroth perturbation order.    
However, in most practical situations, it has ergodicity 
issues beyond the weak-coupling regime. In strongly interacting 
systems, segments of different flavors are typically anti-correlated
so insertion method leads to inefficient importance sampling worsening 
exponentially with $U(\tau''-\tau')$.

The best performance is delivered when we use a combination of the 
two approaches. At zeroth order, we use the insertion measurement since
in this case ergodicity is assured by the probability
factor in (\ref{def_I_0}). For higher-order $G$-configurations we use 
the RS method. With this combination, there is no instability and 
poor statistics can occur only in the unlikely situation with the 
perturbation order sharply peaked around one. Further, when we refer
to the RS method, this combination is implicitly assumed.

\subsection{Selfenergy}
Recently, an efficient estimator for the selfenergy in CT-HYB was proposed
in \cite{Hafermann12}. The method is based on the observation that the selfenergy can be 
expressed from the equation of motion for the Green's function as
\begin{equation}
 \Sigma_\alpha(i\omega_n) = \frac{F_\alpha(i\omega_n)}{G_\alpha(i\omega_n)}
 \label{sigma_estimator_improved}
\end{equation}
where function $F_\alpha(i\omega_n)$ is the Fourier transform of the correlation function
\begin{equation}
 F_{\alpha}(\tau'-\tau'') = 
- \sum_{\beta \neq \alpha} U_{\alpha\beta} 
\langle T\, d^{\phantom \dagger}_\alpha(\tau') d^\dagger_\alpha(\tau'') n_\beta(\tau'') \rangle.
\end{equation}
This method is superior to direct use of the Dyson equation
since the selfenergy is calculated as a ratio of two quantities and only 
relative errors propagate. 
It can be naturally combined with the present measurement procedure. 
All we need to do in order to use estimator (\ref{sigma_estimator_improved})
is to measure the correlation function $F_{\alpha}(\tau'-\tau'')$.
From the definition of the effective time-dependent atomic energy (\ref{effective_atomic_energy})
we see that 
\begin{equation}
 \sum_{\beta\neq \alpha} U_{\alpha\beta} \, n_\beta(\tau;\boldsymbol{\tau}) = E^d_\alpha(\tau;\boldsymbol{\tau}) - E^d_\alpha
\end{equation}
Therefore, regardless of what estimator $G_{\alpha}(\tau', \tau'';\boldsymbol{\tau})$ 
we use for the Green's function, the estimator for the $F$-function is given simply by 
\begin{equation}
 F_{\alpha}(\tau', \tau'';\boldsymbol{\tau}) = G_{\alpha}(\tau', \tau'';\boldsymbol{\tau}) 
 (E^d_\alpha(\tau';\boldsymbol{\tau}) - E^d_\alpha).
\end{equation}
In comparison to measurement of the Green's function itself, this step 
is numerically cheap and requires only an acceptable overhead.

\subsubsection{Dynamic part of the selfenergy}
The selfenergy can be analytically continued to the real axis using the maximum-entropy method \cite{Wang09,Jarrell96} 
in the similar way as the Green's function. 
However, for the MAX-ENT method it is necessary that the continued function 
converges to zero for large frequencies. The full selfenergy does not fulfill this condition
because of the constant Hartree-Fock term. Nevertheless, its dynamic part 
\begin{equation}
\Sigma^D_\alpha(i\omega_n) = \Sigma_\alpha(i\omega_n) - \Sigma^{\rm HF}_\alpha 
\end{equation}
does. So, in order to use the MAX-ENT analytical continuation, we have to measure the Hartree-Fock selfenergy separately 
and at the end of the simulation, we subtract it from the full selfenergy given by equation 
(\ref{sigma_estimator_improved}). We can further reduce error bars of the dynamic part
when the estimator for $\Sigma^{\rm HF}_\alpha$ is chosen so that it is correlated with asymptotic 
behavior of the full selfenergy. This is achieved by separate accumulating of the asymptotic leading
order coefficients $G_\alpha^\infty$
and $F_\alpha^\infty$ and using their ratio as the estimator for $\Sigma^{\rm HF}_\alpha$
\begin{equation}
 \Sigma^D_\alpha(i\omega_n) = \frac{F_\alpha(i\omega_n)}{G_\alpha(i\omega_n)}
 - \frac{F_\alpha^\infty}{G_\alpha^\infty}.
\label{sigma_dynamic_improved}
\end{equation}
The values of $G_\alpha^\infty$ and $F_\alpha^\infty$ depend only
on the connected part of the Green's function and the $F$ function
\begin{align}
 G_\alpha^\infty &= \frac 1 \beta \int_0^\beta d\tau\, G^C_\alpha(\tau_+,\tau)-G^C_\alpha(\tau_-,\tau),
 \nonumber \\
 F_\alpha^\infty &= \frac 1 \beta \int_0^\beta d\tau\,F^C_\alpha(\tau_+,\tau)-F^C_\alpha(\tau_-,\tau).
\end{align}
For the insertion method, these equations reduce to simple expressions 
$G_\alpha^\infty=1$ and $F_\alpha^\infty = \sum_{\beta \neq \alpha} U_{\alpha\beta} \langle n_\beta \rangle$
while for the remove-and-shift approach no such simplifications arise. 

\section{Implementation notes}
\label{section_implementation_notes}
In this section we summarize all steps necessary for implementation
of the algorithm described above. We can provide full source code of 
our implementation upon email request.

First, at the beginning of the simulation, 
in order to avoid repeated computations of eigenvalues of the atomic 
Hamiltonian $E_{loc}$, we can store them into an array indexed by integer representing
the local state (individual bits of the index can naturally represent
occupation numbers of different flavors). 

Next, given a $Z$ configuration from the Metropolis random walk,
we must 
divide the imaginary-time axis to regions with fixed atomic states. 
Each region is bounded by two hybridization events at
$\tau_k$ and $\tau_{k+1}$ and has length $\Delta \tau_k = \tau_{k+1}-\tau_k$.
For each $\tau_k$ and each Matsubara frequency we compute 
$\exp(i\omega_n \tau_k)$ and store it (the factorization property of exponential must 
be used in order to carry out this step efficiently). 

Further, for each flavor $\alpha$ 
we calculate
the effective local atomic energy $E^d_{\alpha k}$ from equation (\ref{effective_atomic_energy}),
exponential factors $\exp(-E^d_{\alpha k} \Delta \tau_k)$
and $d_\alpha(\tau_k;\boldsymbol{\tau}) = \Pi_{l=0}^{k-1} \exp(-E^d_{\alpha l} \Delta \tau_l)$
defined by equation
(\ref{def_dtau}).

For each Matsubara frequency and each region (using the precomputed arrays), we calculate
the following four quantities 
\begin{align}
 a^\alpha_k(i\omega_n) &= \int_{\tau_k}^{\tau_{k+1}} d\tau \exp(-E^d_{\alpha k} (\tau - \tau_k)) \exp(i\omega_n \tau) = 
\nonumber \\
 &(\exp(-E^d_{\alpha k} \Delta \tau_k) \exp(i\omega_n \tau_{k+1}) - \exp(i\omega_n \tau_k))/
 (i\omega_n-E^d_{\alpha k}),
\nonumber \\
 c^\alpha_k(i\omega_n) &= \int_{\tau_k}^{\tau_{k+1}} d\tau \exp(E^d_{\alpha k} (\tau - \tau_k)) \exp(-i\omega_n \tau) = 
\nonumber \\
 &(\exp(-i\omega_n \tau_k) - \exp(E^d_{\alpha k} \Delta\tau_k) \exp(-i\omega_n \tau_{k+1}))/
 (i\omega_n-E^d_{\alpha k}),
 \nonumber \\
 p^\alpha_k(i\omega_n) &= \frac 1 {\beta^2} \int_{\tau^k}^{\tau_{k+1}} d\tau' 
 \int_{\tau'}^{\tau_{k+1}} d\tau''
 \exp(E^d_{\alpha k} (\tau'-\tau'')) \exp(i\omega_n (\tau''-\tau')), \nonumber \\
 &(\exp((i\omega_n-E^d_{\alpha k})\Delta\tau_k) -1 - (i\omega_n-E^d_{\alpha k})\Delta\tau_k)/
 (i\omega_n-E^d_{\alpha k})^2 = 
 \nonumber \\
 &(a^\alpha_k(i\omega_n) \exp(-i\omega_n \tau_k)-\Delta\tau_k)/(i\omega_n-E^d_{\alpha k}),
 \nonumber \\
 h^\alpha_k(i\omega_n) &= - \frac 1 {\beta^2} \int_{\tau^k}^{\tau_{k+1}} d\tau' \int_{\tau_k}^{\tau'} d\tau'' 
 \exp(E^d_{\alpha k} (\tau'-\tau'')) \exp(i\omega_n (\tau''-\tau'))
 = \nonumber \\
 &(1 - \exp(-(i\omega_n-E^d_{\alpha k})\Delta\tau_k) - (i\omega_n-E^d_{\alpha k})\Delta\tau_k)/
 (i\omega_n-E^d_{\alpha k})^2 =
  \nonumber \\
 = &(c^\alpha_k(i\omega_n) \exp(i\omega_n \tau_k)-\Delta\tau_k)/(i\omega_n-E^d_{\alpha k}).
 \label{def_acph}
\end{align}
From these quantities we build-up the functions 
$A^\alpha_j(i\omega_n),\ C^\alpha_j(i\omega_n),\ P^\alpha_j(i\omega_n)$ 
and $H^\alpha_j(i\omega_n)$ 
defined by equations (\ref{def_AC_omega}) and (\ref{def_P_H_omega}). 
In order to use the weight given by equations
(\ref{normalization_weight_separated}) and (\ref{normalization_weight_connected}), we also need to evaluate the corresponding normalization
integrals that can be expressed in terms of zero-frequency limits 
$a^\alpha_j(0_+),\ c^\alpha_j(0_+),\ p^\alpha_j(0_+)$ and $h^\alpha_j(0_+)$. 
Unless $|E^d_{\alpha k}\Delta\tau_k| \ll 1$, we can calculate these 
values simply by putting $i\omega_n=0$ in equations (\ref{def_acph}).
Otherwise, we must use the Taylor expansions
\begin{align}
 a^\alpha_k(0) &= \Delta \tau_k - \frac 1 2 E^d_{\alpha k}\Delta \tau_k^2 +\ldots
 \nonumber \\
 c^\alpha_k(0) &= \Delta \tau_k + \frac 1 2 E^d_{\alpha k}\Delta \tau_k^2 +\ldots
 \nonumber \\
 p^\alpha_k(0) &= \frac{\Delta \tau_k^2}{2} - \frac 1 6 E^d_{\alpha k}\Delta \tau_k^2+\ldots
 \nonumber \\
 h^\alpha_k(0) &= \frac{\Delta \tau_k^2}{2} + \frac 1 6 E^d_{\alpha k}\Delta \tau_k^2+\ldots 
 \label{def_acph0_taylor}
\end{align}
Finally, we have
\begin{align}
  \tilde{A}^\alpha_i(i\omega_n) &=
\sum_{k=k_s}^{k_e} d_\alpha(\tau_k;\boldsymbol{\tau})a^\alpha_k(i\omega_n),
\ {\rm where:}\ \tau_{k_s} = \tau^s_i,\ \tau_{k_e+1} = \tau^s_{i+1},
\nonumber \\
A^\alpha_i(i\omega_n) &= \frac 1 \beta 
\frac{\tilde{A}^\alpha_i(i\omega_n)}{\tilde{A}^\alpha_i(0_+)},
\nonumber \\
\tilde{C}^\alpha_i(i\omega_n) &= \frac 1 \beta 
\sum_{k=k_s}^{k_e} d^{-1}_\alpha(\tau_k;\boldsymbol{\tau})c^\alpha_k(i\omega_n),
\ {\rm where:}\ \tau_{k_s} = \tau^e_{i-1},\ \tau_{k_e+1} = \tau^e_{i},
\nonumber \\
C^\alpha_i(i\omega_n) &= \frac 1 \beta 
\frac{\tilde{C}^\alpha_i(i\omega_n)}{\tilde{C}^\alpha_i(0_+)},
\nonumber \\
\tilde{P}^\alpha_i(i\omega_n) &= 
\sum_{k=k_s}^{k_e} p^\alpha_k(i\omega_n)+
\sum_{k=k_s}^{k_e-1} \sum_{l=k+1}^{k_e}
\frac{d_\alpha(\tau_{l};\boldsymbol{\tau})}{d_\alpha(\tau_{k};\boldsymbol{\tau})}
c^\alpha_k(i\omega_n) a^\alpha_l(i\omega_n),
\nonumber \\
&{\rm where:}\ \tau_{k_s} = \tau^e_{i-1},\ \tau_{k_e+1} = \tau^s_{i+1},
\nonumber \\
P^\alpha_i(i\omega_n) &= \frac 1 {\beta^2}
\frac{\tilde{P}^\alpha_i(i\omega_n)}{\tilde{P}^\alpha_i(0_+)},
\nonumber \\
\tilde{H}^\alpha_i(i\omega_n) &= 
\sum_{k=k_s}^{k_e} h^\alpha_k(i\omega_n)+
 \sum_{k=k_s}^{k_e-1} \sum_{l=k+1}^{k_e}
\frac{d_\alpha(\tau_k;\boldsymbol{\tau})}{d_\alpha(\tau_l;\boldsymbol{\tau})}
a^\alpha_k(i\omega_n) c^\alpha_l(i\omega_n),
\nonumber \\
&{\rm where:}\ \tau_{k_s} = \tau^s_i,\ \tau_{k_e+1} = \tau^e_{i+1},
\nonumber \\
H^\alpha_i(i\omega_n) &= \frac 1 {\beta^2}
\frac{\tilde{H}^\alpha_i(i\omega_n)}{\tilde{H}^\alpha_i(0_+)}.
\label{def_ACPH_from_acph}
\end{align}
Evaluation of the double sums in definitions of $\tilde{P}^\alpha_i(i\omega_n)$ 
and $\tilde{H}^\alpha_i(i\omega_n)$ can be accelerated when results from 
evaluation of $\tilde{A}^\alpha_i(i\omega_n)$ and $\tilde{C}^\alpha_i(i\omega_n)$ 
are reused and its cost is linear in number of covered regions $k_e-k_s$.
The zeroth order term (\ref{def_I_0}) can be evaluated analogously, but 
order in which individual summations are carried out must be chosen based on
sign of $d_\alpha(\beta;\boldsymbol{\tau})-1$ in order to avoid the numerical 
instability. Finally, with results of equations (\ref{def_ACPH_from_acph}),
we can compute estimators (\ref{Gw_separated}) and (\ref{Gw_conn_RS}).

\subsection{Computational complexity}
The obvious disadvantage of the algorithm described in this paper compared
to the standard measurement is the increased computational cost. 
Unlike the standard approach, our measurement is causing a non-negligible slow-down
of the entire simulation. 

Overall, the computational complexity of the improved measurement scales 
\emph{quadratically} with the perturbation order and linearly with the number 
of measured Matsubara frequencies because of evaluation of equation (\ref{Gw_separated}).
Nevertheless, the standard estimator (\ref{G_estimator_standard}) 
has a similar form and it is 
not the real computational bottleneck for any reasonable perturbation order. 
In practice, the most expensive part of our algorithm is computation of
the functions $A^\alpha_j(i\omega_n)$, $C^\alpha_j(i\omega_n)$, $P^\alpha_j(i\omega_n)$ 
and $H^\alpha_j(i\omega_n)$ from equation (\ref{def_ACPH_from_acph})
(including all of its inputs). Its cost, however, is only
\emph{linear} in the total number of regions, Matsubara frequencies
and flavors. Therefore, relative cost of measurement compared to 
one MC step is decreasing with increasing perturbation order (temperature).
Nevertheless, it is essential not to perform 
measurement after every Monte-Carlo step but only 
approximately once the autocorrelation time to keep the 
simulation speed reasonable.

\section{Results}
\label{section_results}
In this section, we present the numerical results 
for two test cases. First, the single band Anderson impurity 
model with the semielliptic density of states (half band-width is set to one).
Second, as an example of a realistic material calculation, we show results for 
the cobalt atom in LaCoO${}_3$. If not explicitly stated otherwise, we use the 
RS measurement. As a benchmark, results obtained with delta estimators
measured to Matsubara frequencies and Legendre polynomials are used.
Our implementation is based on the hybridization-expansion solver \cite{Gull11b} 
from the ALPS library version 2 \cite{Bauer11,Albuquerque07}.  
All the other results shown here 
were also obtained using the ALPS library.

\begin{figure}
\centering
  \includegraphics[width=6cm]{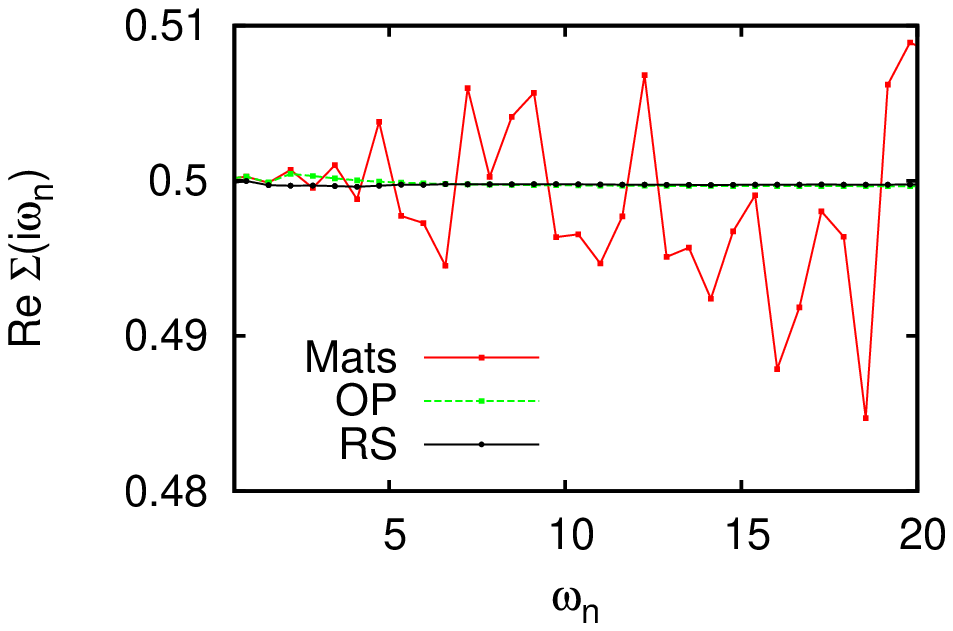} \includegraphics[width=6cm]{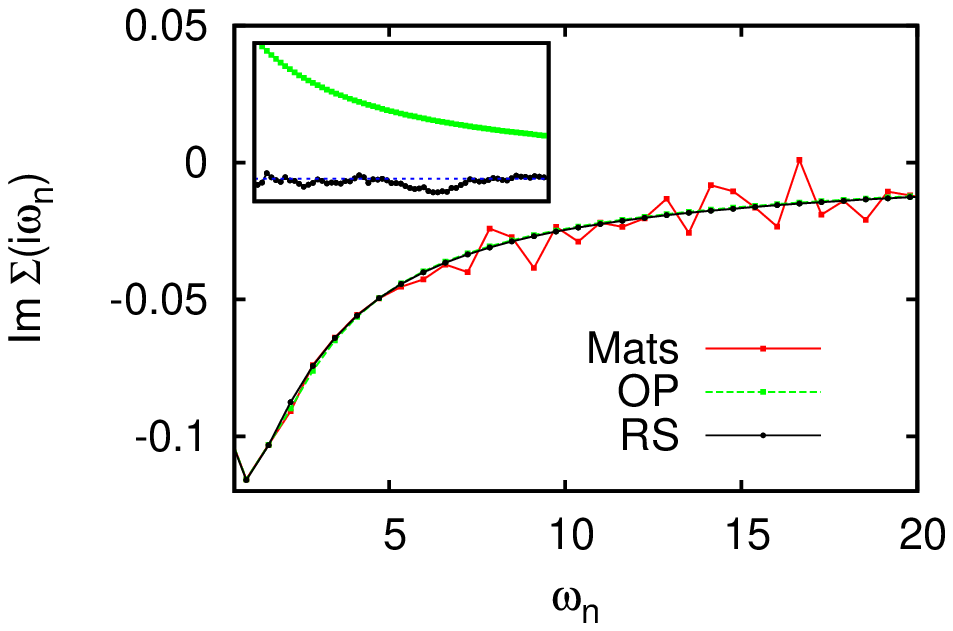}
 \caption{The real and the imaginary part of the selfenergy for $U=1$ and $\beta=10$. The red and the green
line were obtained from delta estimators measured into Matsubara frequencies and Legendre polynomials. 
Inset in the right graph shows the difference between the measured selfenergy and the analytically computed
high-frequency tail.}
\label{graph_Bethe_im_re_sigma}
\end{figure}

In figure \ref{graph_Bethe_im_re_sigma}, the impurity selfenergy for the half-filled Anderson model for 
$U=1$ and $\beta=10$ (mean perturbation order 1.8) is plotted. 
The results were obtained from the 
equation of motion (\ref{sigma_estimator_improved}) using independent simulations
and the same amount of CPU time.
We intentionally use 
short MC runs here in order to make the statistical noise visible to the naked eye. 
While noise from the standard method is increasing with frequency, no
such effect is visible when continuous estimators or measurement into orthogonal polynomials 
are used. 
With the orthogonal polynomials, however, the stochastic noise is suppressed
at the price of introducing strong statistical correlations
between the values at different Matsubara frequencies. 
Correlations in the imaginary part of the selfenergy obtained from the RS method are much weaker.
This is demonstrated in the inset which shows the difference between the measured selfenergy
and the analytically computed high-frequency tail. 
Errors of the real part of the selfenergy are strongly correlated
because of uncertainty in the Hartree-Fock term. 
These correlations can be eliminated in the dynamic selfenergy by means of equation 
(\ref{sigma_dynamic_improved}). 

\begin{figure}
\centering
  \includegraphics[width=6cm]{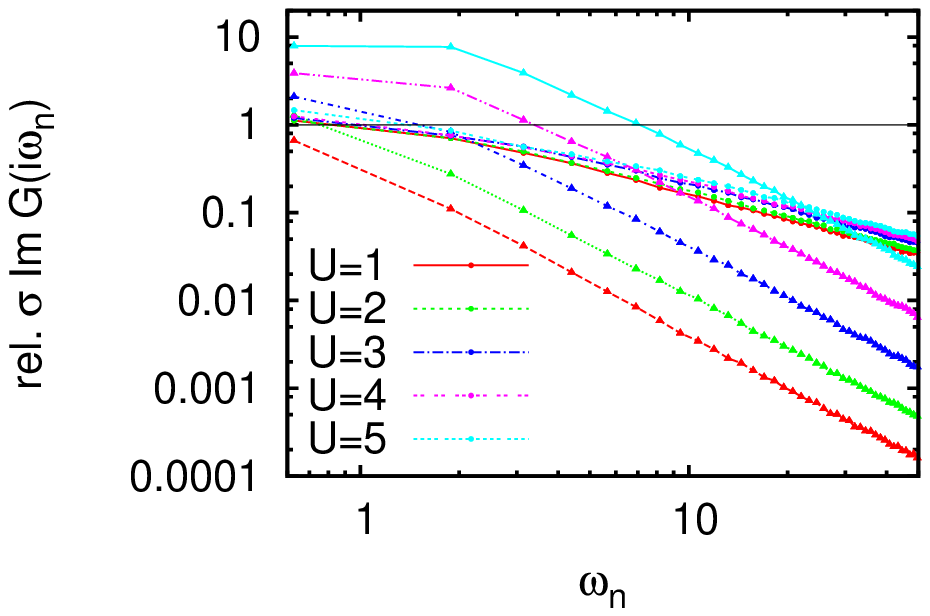} 
  \includegraphics[width=6cm]{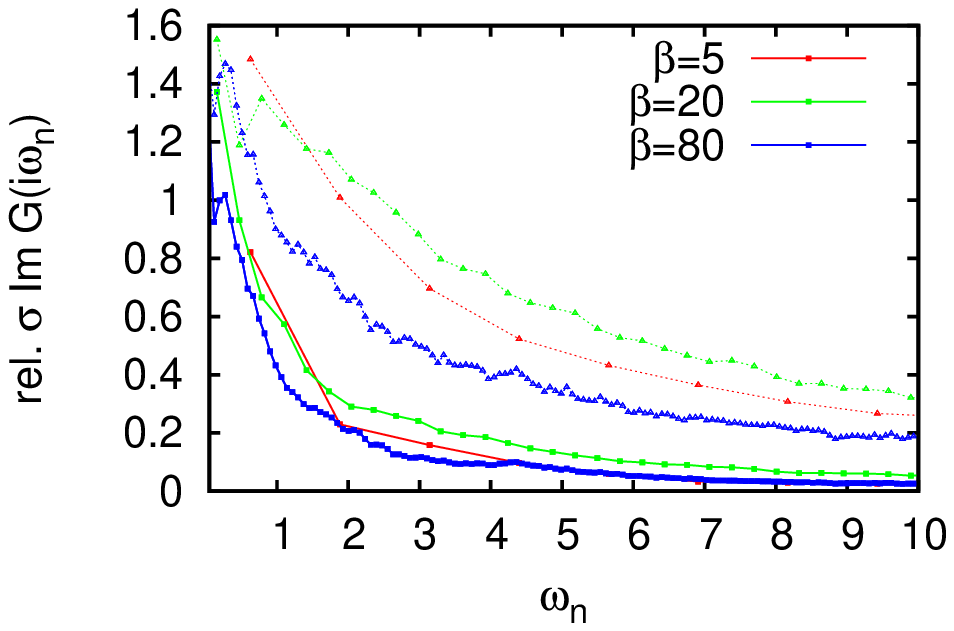} 
 \caption{Left: Comparison of relative error bars of the imaginary part of the Green's function
 for $U=1,2,3,4$ and $5$ and $\beta=5$.
 Errors from the insertion approach (labeled with triangles) are lower in the weak-coupling regime, 
 but they grow exponentially with increasing interaction. On the contrary, the RS 
 measurement (labeled with circles) yields almost interaction-independent results. 
 Right: Relative errors for $U=1$ and different temperatures with (squares) and without (triangles)
 the correction on the correct asymptotic behavior. 
 The mean perturbation order here grows approximately linearly with temperature 
 up to $\langle K \rangle \approx 16$ at $\beta=80$.
}
\label{graph_errors_RSI_Bethe}
\end{figure}

Next, we show comparison of accuracies of different methods in figure \ref{graph_errors_RSI_Bethe}.
We plot the relative error, that is, the ratio of the errors obtained with continuous estimators
and the error from the standard formula. 
The comparison between the insertion method and the RS method shows that while the first one has better
high-frequency behavior, it is (exponentially) unstable in the strong coupling regime. On the contrary, 
the RS method yields more accurate results regardless of $U$. The temperature dependence of the relative 
error from the RS measurement shows that benefits of this approach persist even in higher perturbation 
orders. The correction on the exact $1/\omega$ asymptotic discussed in section \ref{section_RS_approach} 
further increases the accuracy. 

\begin{figure}
\begin{tabular}{cc}
 \includegraphics[width=6cm]{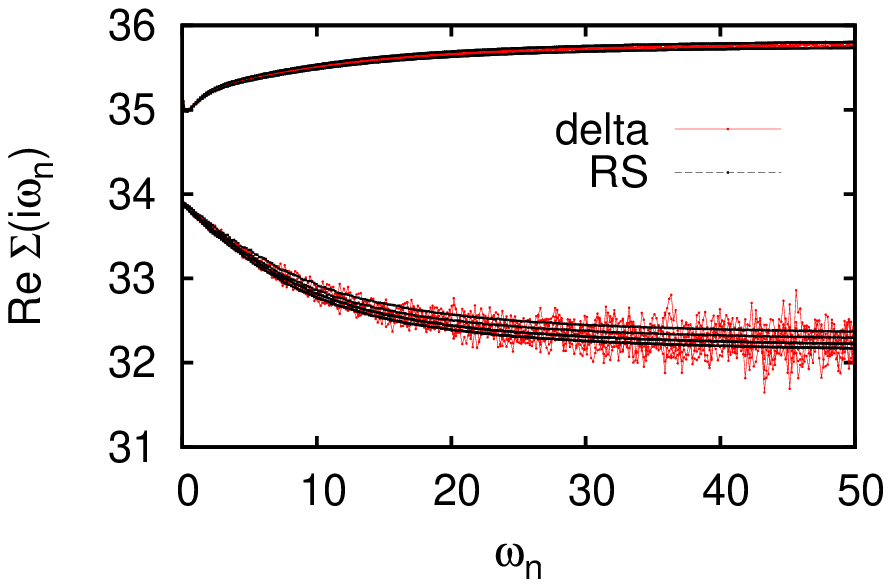} 
 &
 \includegraphics[width=6cm]{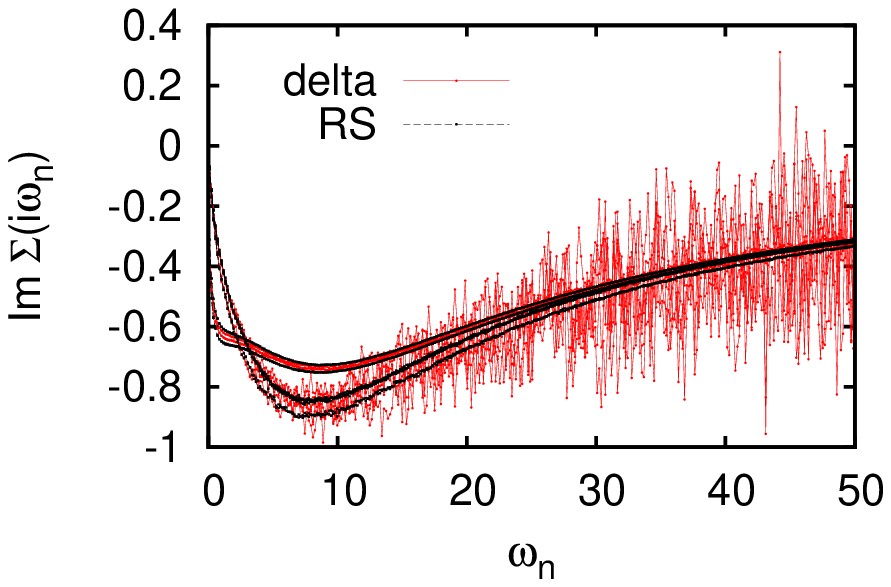} 
 \\
 \includegraphics[width=6cm]{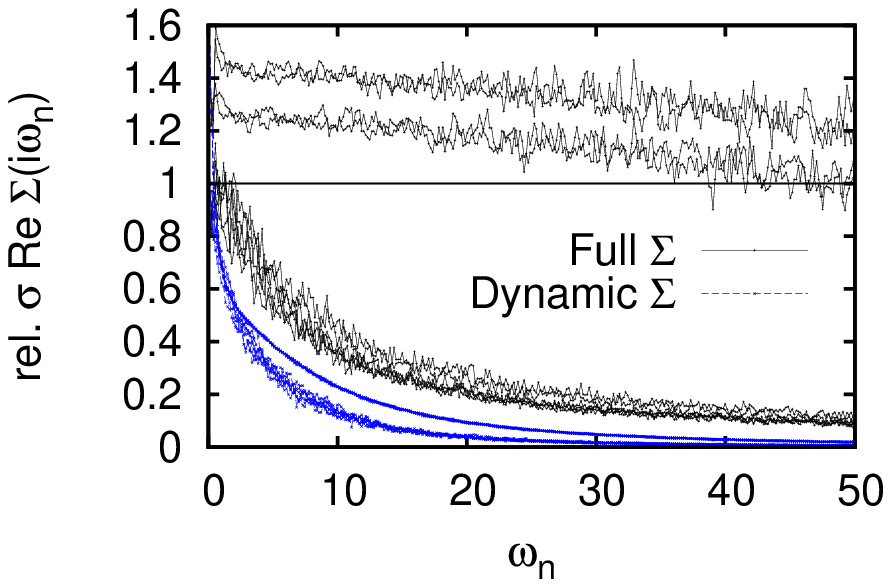} 
 &
 \includegraphics[width=6cm]{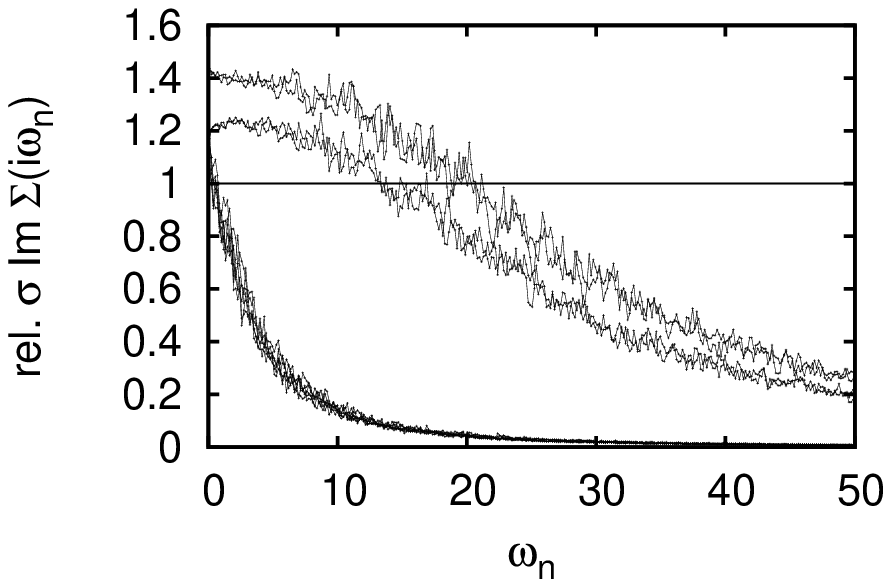} 
\end{tabular}
\caption{Top: The selfenergy of lanthanum cobaltite calculated with delta estimators and the RS method. 
The two branches correspond to the weakly hybridized $t_{2g}$ orbitals and the strongly hybridized $e_g$ orbitals. 
Bottom: Relative errors of the full selfenergy and its dynamic part.}
\label{figure_sigma_lacoo3}
\end{figure}

In figure \ref{figure_sigma_lacoo3}, the selfenergy of 
lanthanum cobaltite is plotted.
The impurity model used describes the full d-shell of cobalt
with five orbitals. The perturbation order is strongly
imbalanced here. While the mean perturbation order of 
$t_{2g}$ orbitals is very low, $\langle K \rangle \approx 0.3$,
$e_g$ orbitals are strongly hybridized with $\langle K \rangle \approx 42$.
The $e_g$ selfenergy measured with the RS method is less accurate than 
the one obtained from the standard formula apart from the high-frequency tail. 
The reason for this is that the same amount of CPU time was used for 
both simulations and only approximately 60\% of 
Monte Carlo steps were done with the slower RS method. 

\section{Conclusions and outlook}
The presented measurement algorithm for the Green's function 
has several possible applications. First of all, it solves the 
poor statistics problem at low perturbation orders. At higher 
orders, it yields stochastic noise decreasing with Matsubara
frequency so it can generate better data for the numerical 
analytical continuation of both the Green's function and the selfenergy.
In principle, it also allows for measurement of a general local susceptibility 
of non density-density type that is very difficult to obtain 
with the standard method.

Its main disadvantage is relatively high computational complexity
resulting in slowdown of the entire simulation and slightly 
worse accuracy at low frequencies.
Moreover, 
since our method essentially relies on fast and simple evaluation of
the local trace, generalization to the matrix formalism would be 
difficult, if possible at all.

\section{Acknowledgement}
We acknowledge inspiring discussions with Marcus Kollar, Liviu Chioncel and Junya Otsuki
during development of this method.
The research on this project was supported by Deutsche Forschungsgemeinschaft Research Unit FOR 1346. 
All calculations were done on the Dorje cluster at the Institute of Physics of the Czech 
Academy of Sciences in Prague. 

\bibliographystyle{elsarticle-num}
\bibliography{ref} 

\end{document}